\documentclass[12pt,epsfig,prd,showpacs]{revtex4}
\usepackage{graphicx}
\usepackage{epsfig}
\usepackage{dcolumn}
\usepackage{bm}
\usepackage{overpic}

\newcommand{\jpsi}{J/\psi}
\newcommand{\psip}{\psi(2S)}

\newcommand{\kkppp}{K^+K^-\pi^+\pi^-\pi^0}

\newcommand{\kstarp}{K^{\ast}(892)^{+}}

\newcommand{\kstaro}{K^{\ast}(892)^0}

\begin{document}
\begin{center}

\centerline{\bf \boldmath \large Experimental study of $\psip$
decays to $\kkppp$ final states} \vspace{0.4cm}
 M.~Ablikim$^{1}$,
J.~Z.~Bai$^{1}$, Y.~Ban$^{12}$, J.~G.~Bian$^{1}$, X.~Cai$^{1}$,
H.~F.~Chen$^{17}$, H.~S.~Chen$^{1}$, H.~X.~Chen$^{1}$,
J.~C.~Chen$^{1}$, Jin~Chen$^{1}$
% ...
\\
\vspace{0.2cm}
(BES Collaboration)\\
\vspace{0.2cm} {\it
$^{1}$ Institute of High Energy Physics, Beijing 100049, People's Republic of China\\
% ...
$^{12}$ Peking University, Beijing 100871, People's Republic of China\\
% ...
$^{17}$ University of Science and Technology of China, Hefei 230026, People's Republic of China\\
% ...
}
\end{center}

\vspace{0.4cm}

\begin{abstract}

$K^+K^-\pi^+\pi^-\pi^0$ final states are studied using a sample of
$14\times10^6$  $\psi(2S)$ decays collected with the Beijing
Spectrometer (BESII) at the Beijing Electron-Positron Collider. The
branching fractions of $\psi(2S)$ decays to $
K^+K^-\pi^+\pi^-\pi^0$, $\omega K^+ K^-$, $\omega f_0(1710)$, $
\kstaro K^- \pi^+\pi^0+c.c.$, $ \kstarp K^- \pi^+\pi^-+c.c.$,
$\kstarp K^- \rho^0+c.c.$ and $\kstaro K^-\rho^+ + c.c.$ are
determined. The first two agree with previous measurements, and the
last five are first measurements.
\end{abstract}
\maketitle \vspace{5mm}

\section{Introduction}
From perturbative QCD (pQCD), it is expected that both $J/\psi$ and
$\psip$ decaying into light hadrons are dominated by the annihilation
of $c \bar c$ into three gluons, with widths proportional to the
square of the wave function at the origin
$|\Psi(0)|^2$~\cite{T.Appelquist}.  This yields the pQCD ``$12\%$''
rule:
$$Q_h = \frac{B_{\psip\to h}}{B_{J/\psi\to h}}\approx\frac{B_{\psip\to
e^+e^-}}{B_{J/\psi\to e^+e^-}}\approx12\% .$$ The violation of the
above rule was first observed in the $\rho \pi$ and $K^{\ast +}K^-
+c.c.$ decay modes by Mark-II~\cite{Mark-II}. Following the scenario
proposed in Ref.~\cite{J.L.Rosner}, that the small $\psip \to \rho
\pi$ branching fraction is due to the cancelation of the S- and
D-wave matrix elements in $\psip$ decays, it was suggested that all
$\psip$ decay channels should be affected by the same S- and D-wave
mixing scheme, and thus all ratios of branching fractions of $\psip$
and $J/\psi$ decays into the same final state could have values
different from $12\%$, expected between pure 1S and 2S
states~\cite{P.Wang}. The mixing scenario also predicts $\psi(3770)$
decay branching fractions since the $\psi(3770)$ is a mixture of S-
and D-wave charmonia, as well.  Many channels of $J/\psi$, $\psip$
decays, and $\psi(3770)$ decays should be measured to test this
scenario.

In this paper, we report measurements of $\kkppp$ final states, as
well as some intermediate states that decay to the same final states.
The data samples used for this analysis consist of
$14.0\times10^6(1\pm 4\%)$ $\psip$ events taken at $\sqrt s =3.686$
GeV~\cite{X.H.Mo} and $6.42(1\pm 4\%)$pb$^{-1}$ of continuum data at
$\sqrt s=3.65$ GeV~\cite{S.P.Chi}.

\section{BESII detector}

BESII is a large solid-angle magnetic spectrometer which is
described in detail in Ref.~\cite{bes2}. The momentum of  charged
particles is determined by a 40-layer cylindrical main drift chamber
(MDC) which has a momentum resolution of
$\sigma_{p}$/p=$1.78\%\sqrt{1+p^2}$ ($p$ in GeV/c).  Particle
identification (PID) is accomplished using specific ionization
($dE/dx$) measurements in the drift chamber and time-of-flight (TOF)
information in a barrel-like array of 48 scintillation counters. The
$dE/dx$ resolution is $\sigma_{dE/dx}\simeq8.0\%$; the TOF
resolution for Bhabha events is $\sigma_{TOF}= 180$ ps.  Radially
outside of the time-of-flight counters is a 12-radiation-length
barrel shower counter (BSC) comprised of gas tubes interleaved with
lead sheets. The BSC measures the energy and direction of photons
with resolutions of $\sigma_{E}/E\simeq21\%/\sqrt{E}$ ($E$ in GeV),
$\sigma_{\phi}=7.9$ mrad, and $\sigma_{z}=2.3$ cm. The iron flux
return of the magnet is instrumented with three double layers of
proportional counters that are used to identify muons.

A GEANT3 based Monte Carlo (MC) simulation package~\cite{simbes},
which simulates the detector response, including interactions of
secondary particles in the detector material, is used to determine
detection efficiencies and mass resolutions, as well as to optimize
selection criteria and estimate backgrounds. Reasonable agreement
between data and MC simulation is observed in various channels
tested, including $e^+e^-\to(\gamma) e^+ e^-$,
$e^+e^-\to(\gamma)\mu^+\mu^-$, $J/\psi\to p\bar{p}$,
$J/\psi\to\rho\pi$, and $\psi(2S)\to\pi^+\pi^- J/\psi$, $J/\psi\to
l^+ l^-$.

\section{Event  Selection}

The $K^+K^-\pi^+\pi^-\pi^0$ final states are reconstructed with four
charged tracks and two photons.

\subsection{Photon and charged particle identification}
A neutral cluster is considered to be a good photon candidate if the
following requirements are satisfied: it is located within the BSC
fiducial region, the energy deposited in the BSC is greater than 50
MeV, the first hit appears in the first six-radiation lengths, the angle
between the cluster development direction in the BSC and the photon
emission direction is less than
$37^\circ$, and the angle between the cluster and the nearest charged
particle is greater than $15^\circ$.

Each charged track is required to be well fit by a three dimensional
helix, to originate from the interaction region,
$V_{xy}=\sqrt{V_x^2+V_y^2}<2.0$ cm and $|V_z|<20$ cm, and to have a
polar angle $|\cos\theta|<0.8$. Here $V_x$, $V_y$, and $V_z$ are the
$x$, $y$, $z$ coordinates of the point of closest approach of the
track to the beam axis. The TOF and $dE/dx$ measurements for each
charged track are used to calculate $\chi_{PID}^2(i)$ values and the
corresponding confidence levels $Prob_{PID}(i)$ for the hypotheses
that a track is a pion, kaon, or proton, where $i~(i=\pi/K/p)$ is the
particle type.

\subsection{Selection criteria}

For the final states of interest, the candidate events are
required to satisfy the following selection criteria:
\begin{enumerate}
\item The number of charged tracks in the MDC equals four with net charge zero.
\item The number of good photon candidates equals two or three.
\item For each charged track, the particle identification confidence
    level for a candidate particle assignment is required to be
    greater than $1\%$.
\item The angle between two photons satisfies $\theta_{\gamma
      \gamma} > 6^\circ$ to remove the background from split-off fake photons.
  \item To reduce contamination
    from $\psip \to \eta J/\psi$ with $\eta \to \pi^+\pi^-\pi^0$ and
    $J/\psi \to \mu^+ \mu^-$, $N_{+}^{hit}+N_{-}^{hit}<4$ is used,
    where $N_{+}^{hit}$ is the number of muon
    identification layers with matched hits for the higher momentum
    positive charged track and ranges from 0 to 3, indicating not a
    muon (0) or a weakly (1), moderately (2), or strongly (3)
    identified muon~\cite{muon}, and $N_{-}^{hit}$ is the
    corresponding number for the higher momentum negative charged
    track.
   \item To reject background from $\psip \to \pi^+\pi^- J/\psi$,
$|M_{recoil}^{\pi^+ \pi^-}-3.097|>0.05$ $\hbox{GeV}/c^2$ is used,
where $M_{recoil}^{\pi^+ \pi^-}$ is the mass recoiling against the
$\pi^+\pi^-$ pair.
\end{enumerate}

To improve track momentum resolution and reduce background, four
constraint kinematic fits imposing energy and momentum conservation
are performed. We loop over all combinations of charged tracks and
good photons and select the one with the minimum $\chi^2_{com}$ for
the assignment $\psi(2S)\to \gamma\gamma \pi^+\pi^- K^+K^-$, where
the combined $\chi^2$, $\chi^2_{com}$, is defined as the sum of the
$\chi^2$ values of the kinematic fit, $\chi^2_{kine}$, and those
from each of the four particle identification assignments:
$\chi^2_{com}=\sum\limits_{i} \chi_{PID}^2(i) + \chi_{kine}^2.$ We
require $\chi^2_{com}(\gamma \gamma \pi^+ \pi^- K^+
K^-)<\chi^2_{com}(\gamma \gamma \pi^+\pi^-\pi^+ \pi^-),$
$\chi^2_{com}(\gamma \gamma \pi^+ \pi^- K^+ K^-)<\chi^2_{com}(\gamma
\gamma K^+ K^- K^+ K^-),$ and the confidence level of the kinematic
fit to be greater than $1\%$.

After applying the above selection criteria, there is still
background remaining from $\psip \to K_S K^\pm \pi^\mp \pi^0$, $K_S \to \pi^+
\pi^-$. This is removed by requiring $m_{\pi^+ \pi^-} > 0.51$
$\hbox{GeV}/c^2$ or $m_{\pi^+ \pi^-} <0.45$ $\hbox{GeV}/c^2$,
$m_{K^+ \pi^-} > 0.51$ $\hbox{GeV}/c^2$ or $m_{K^+ \pi^-} < 0.47$
$\hbox{GeV}/c^2$, and $m_{K^- \pi^+} > 0.51$ $\hbox{GeV}/c^2$ or $m_{K^-
\pi^+} < 0.47$ $\hbox{GeV}/c^2$, where in calculating $m_{K\pi}$, the
 $\pi$ mass is used for the $K^{\pm}$ track.

\section{Data analysis}

\begin{figure}[htbp]
\centerline{\psfig{file=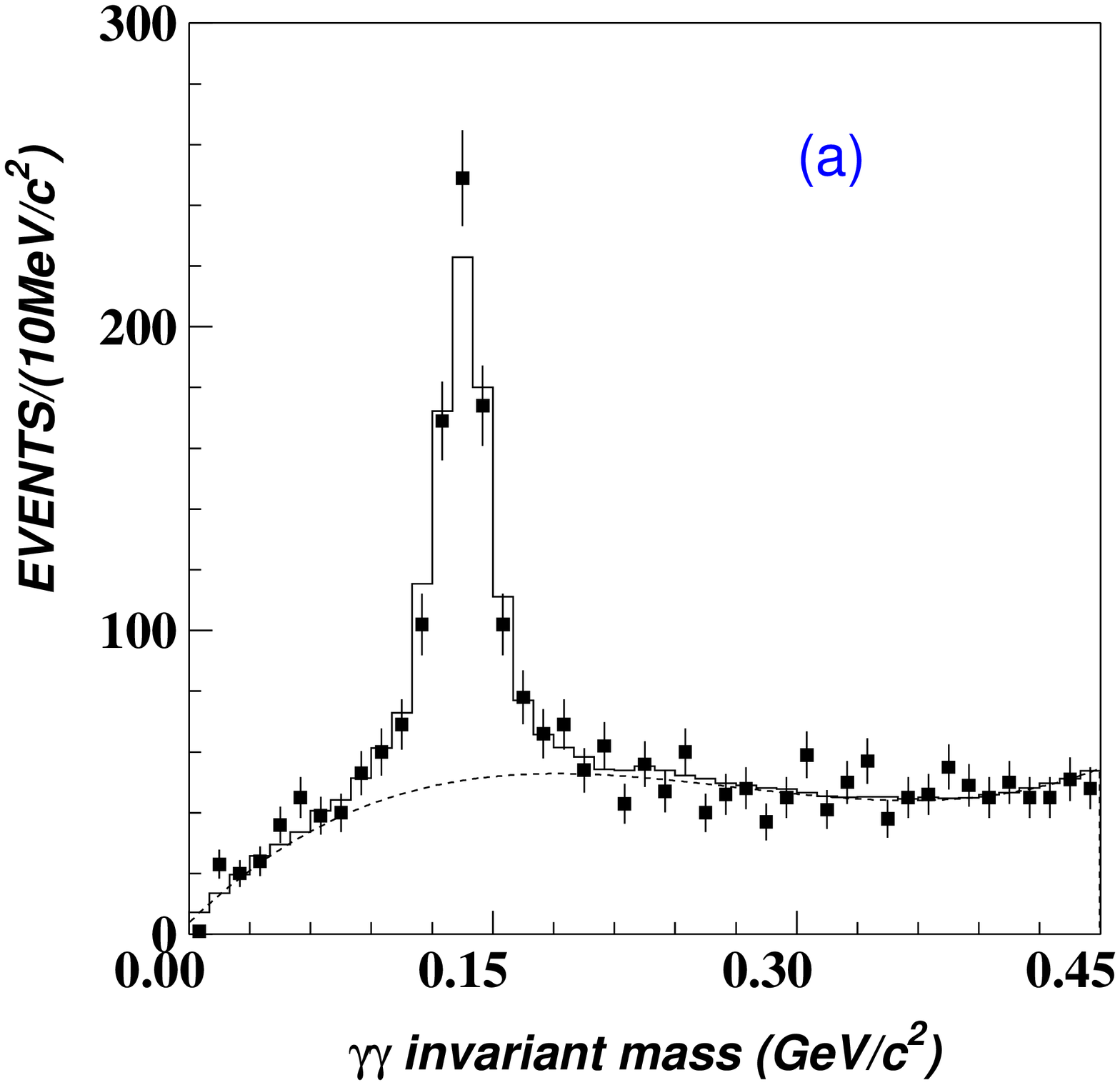,width=7cm,height=7cm}
            \psfig{file=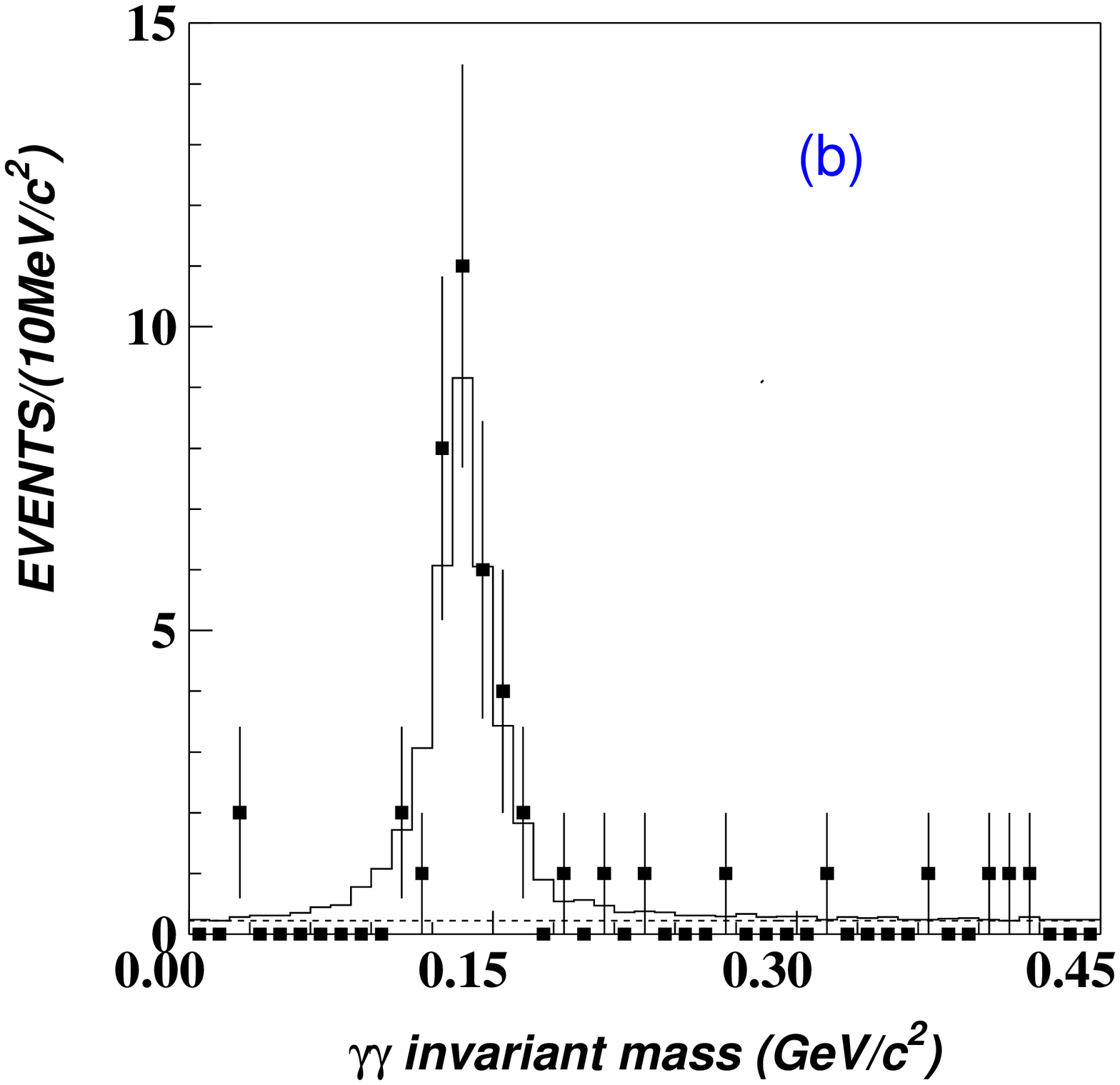,width=7cm,height=7cm}}
\caption{\label{2k3pi} The $\gamma \gamma$ invariant mass
distributions of (a) selected $\psip\to K^+K^-\pi^+\pi^-\gamma \gamma$
candidate events and (b) selected $e^+e^-\to
K^+K^-\pi^+\pi^-\gamma \gamma$ candidate events at $\sqrt{s}=3.65$ GeV. The
squares with error bars are data, the histograms are the fit, and
the dashed curves are background shapes from the fit.}
\end{figure}

The $\gamma \gamma$ invariant mass distribution for events that
survive the selection criteria is shown in Fig.~\ref{2k3pi}(a), where
a clear $\pi^0$ signal can be seen.  By fitting this distribution with
a $\pi^0$ signal shape, obtained from MC simulation, and a 3rd order
background polynomial, $698 \pm 41$ events are obtained in the $\sqrt
s=3.686$ GeV data sample. From exclusive Monte Carlo simulation, we
determine that the main background channels are from $\psip \to \pi^0
\pi^0 \jpsi, \jpsi \to K^+ K^- \pi^+ \pi^-$, $\psip \to \gamma
\chi_{cJ}, \chi_{cJ}\to K^+ K^- \pi^+ \pi^-$, and $\psip \to \gamma
\chi_{cJ}, \chi_{cJ} \to \kkppp$. However, the $\gamma \gamma$
invariant mass distributions from all these background
channels do not have a peak at $m_{\gamma \gamma} = m_{\pi^0}$, and,
therefore, they will not contribute to the number of fitted events.
Applying the same selection criteria to the $\sqrt{s}=3.65$ GeV data,
we obtain the $\gamma \gamma$ invariant mass distribution shown in
Fig.~\ref{2k3pi}(b). Fitting Fig.~\ref{2k3pi}(b) in a similar way as
Fig.~\ref{2k3pi}(a) yields $35 \pm 7$ events. The efficiencies are
$(3.68\pm0.05)\%$ for $\psip \to K^+ K^- \pi^+\pi^-\pi^0$ and
$(2.59\pm0.05)\%$ for continuum $e^+e^-\to K^+K^-\pi^+\pi^-\pi^0$,
where the difference is due to the initial state radiation for
continuum data. Here in calculating efficiencies, we have considered
the contributions from important intermediate states (shown later in
the paper) in $ \psip \to K^+ K^- \pi^+\pi^-\pi^0$.

\begin{figure}[htbp]
\centerline{\psfig{file=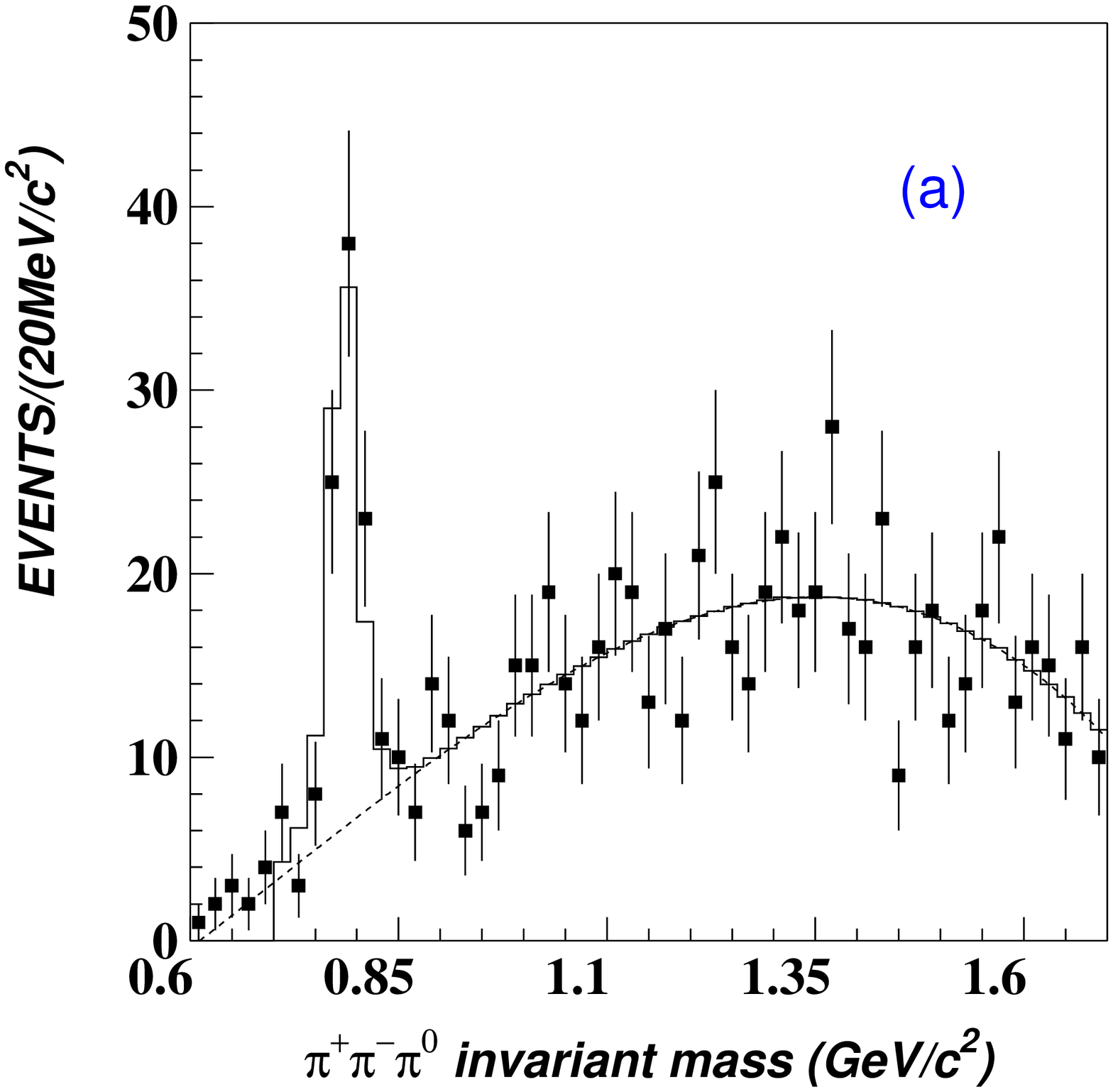,width=7cm,height=7cm}
            \psfig{file=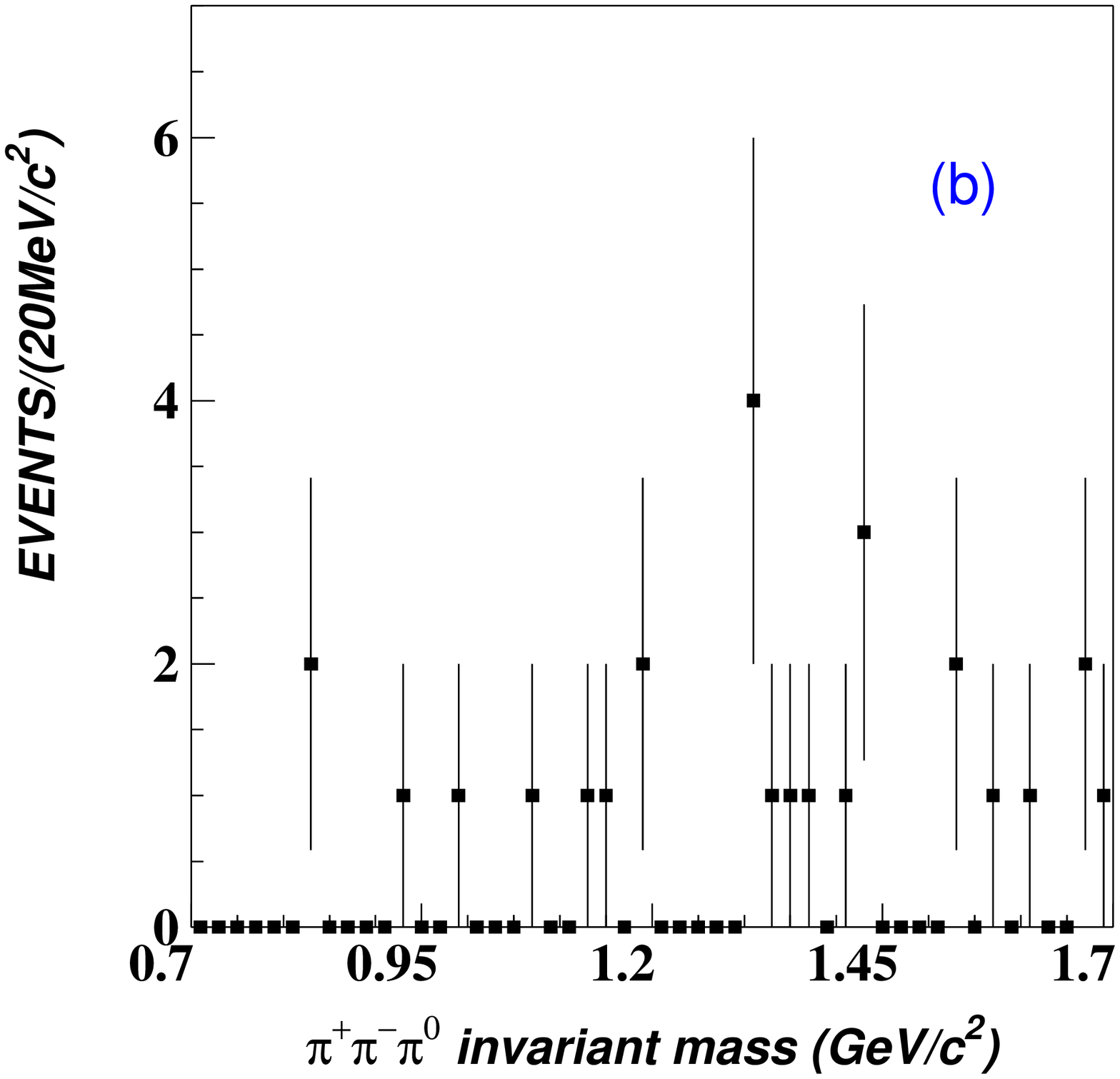,width=7cm,height=7cm}}
        \caption{\label{wkk-fit} The $\pi^+\pi^-\pi^0$ invariant mass
distributions of (a) selected $\psip \to K^+K^-\pi^+\pi^-\pi^0 $
candidate events and (b) selected $e^+e^-\to K^+K^-\pi^+\pi^-\pi^0$
candidate events at $\sqrt{s}=3.65$ GeV. The squares with error bars
are data, the histograms are the fit, and the dashed curves are
background shapes from the fit.}
\end{figure}

In the following analysis, we study the $\pi^+ \pi^- \pi^0$, $K^+
K^-$, $K^+ \pi^- +c.c.$, $K^+ \pi^0 +c.c.$ and $\pi^+ \pi^-$
invariant mass spectra to look for possible intermediate resonance
states. Fig.~\ref{wkk-fit}(a) shows the $\pi^+\pi^-\pi^0$ invariant
mass distribution for $K^+ K^- \pi^+ \pi^- \gamma \gamma$ events
after requiring $|m_{\gamma \gamma}-0.135|<0.03$ $\hbox{GeV}/c^2$
(to increase the efficiency, the requirements on $m_{\pi^+ \pi^-}$
and $m_{K \pi}$ used for the $K_S$ veto are removed); a clear
$\omega$ signal is observed. A fit  with an $\omega$ signal shape
obtained from MC simulation and a polynomial background gives $78
\pm 11 $ signal events with a statistical significance of
8.0$\sigma$. The detection efficiency for this decay mode is $(
2.66\pm 0.04)\%$, where we have considered the contributions from
important intermediate states. Fig.~\ref{wkk-fit}(b) shows the
$\pi^+ \pi^- \pi^0$ invariant mass distribution for $\sqrt{s}=3.65$
GeV data, where we obtain $0^{+1.3}_{-0}$ $\omega K^+K^-$ events at
the $68.3\%$ confidence level with a similar fit to the
$\pi^+\pi^-\pi^0$ invariant mass spectrum.

\begin{figure}[htbp]
 \centerline{\psfig{file=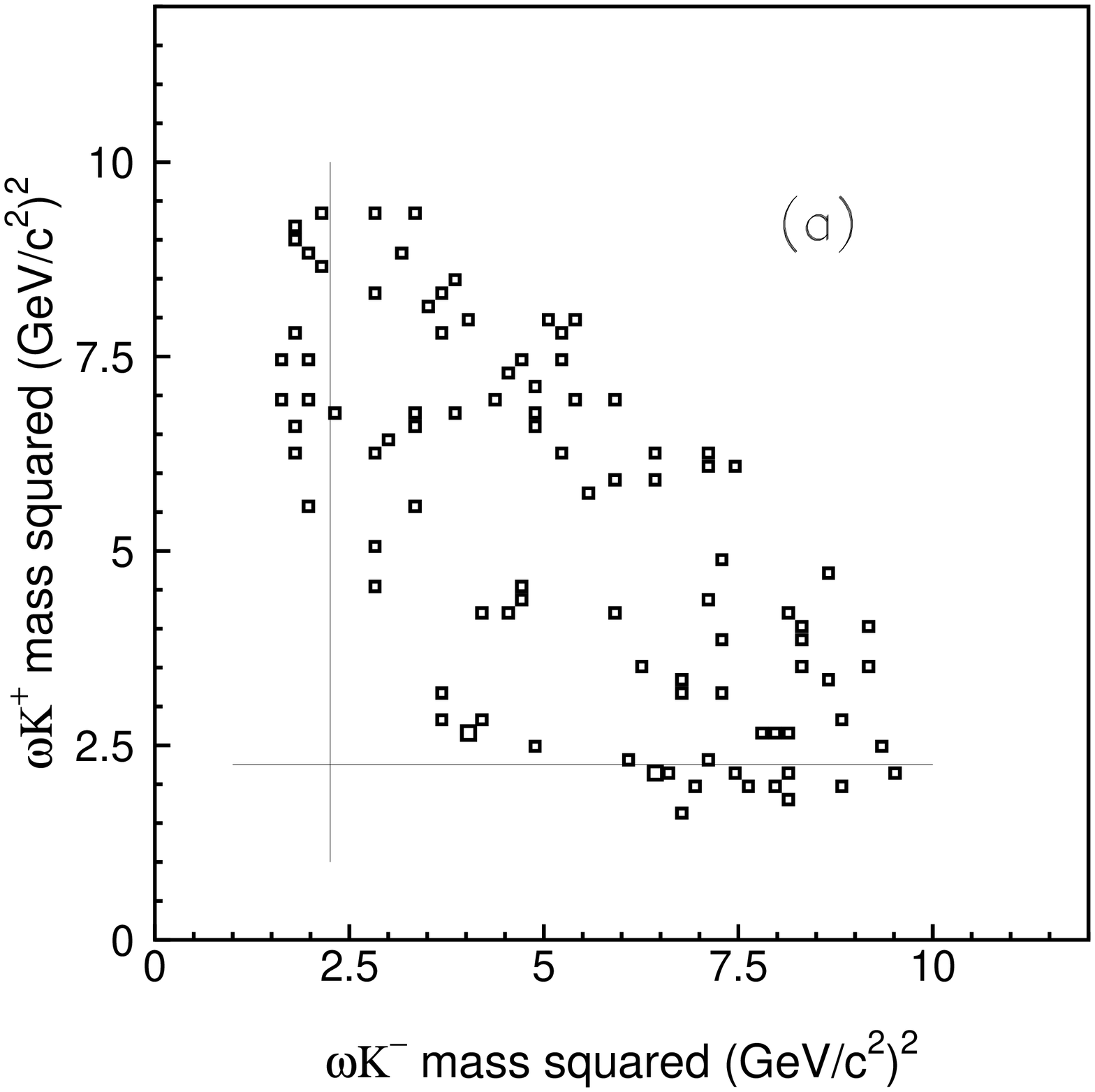,width=7cm,height=7cm}
             \psfig{file=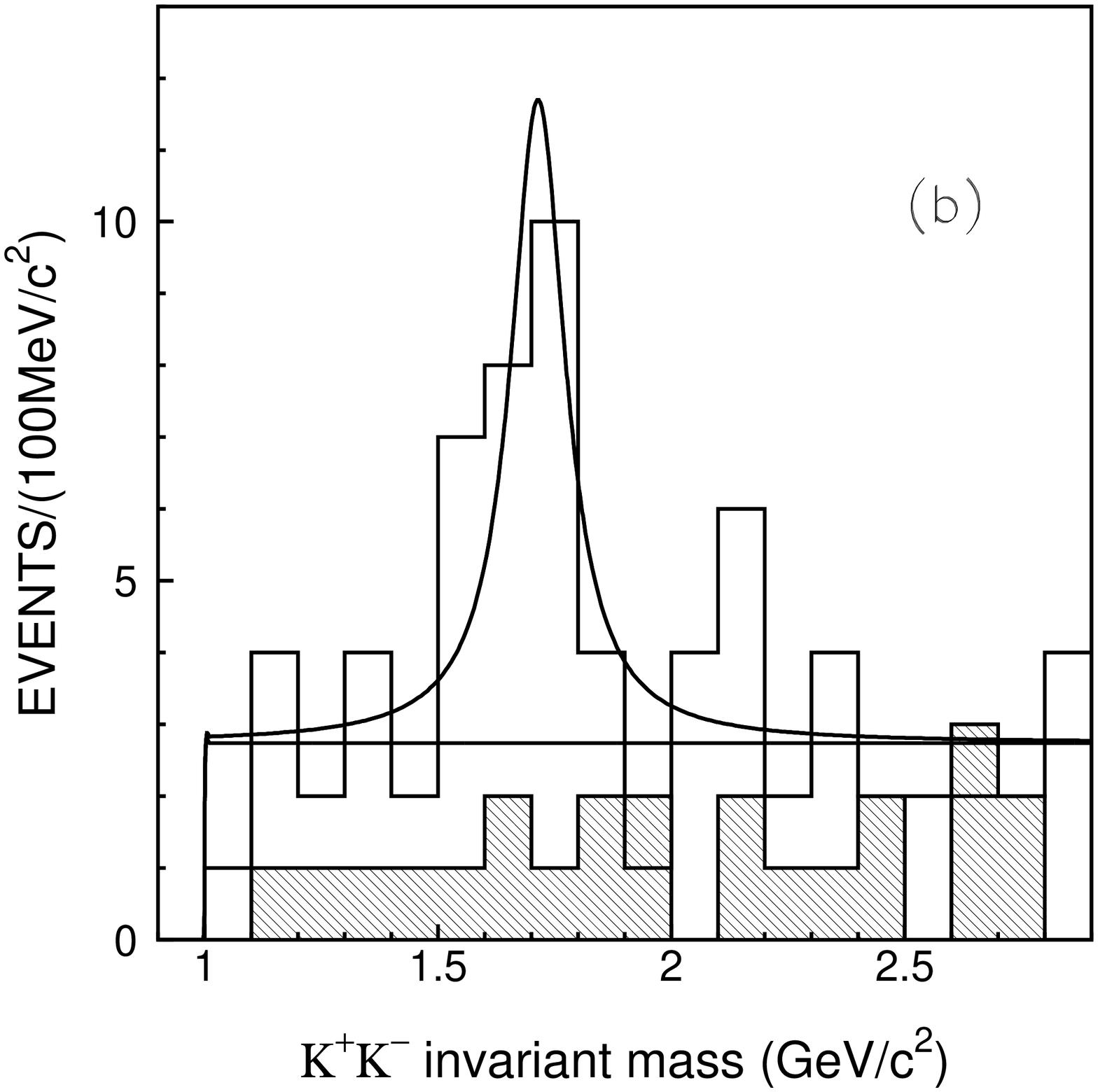,width=7cm,height=7cm}}
         \caption{\label{mkaka} (a) The Dalitz plot of selected $\psip
\to \omega K^+ K^-$ candidates, and (b) the $K^+ K^-$ invariant mass
distribution for candidate $\psip \to \omega K^+ K^-$ events after the
$K_1(1270)$ veto ($m_{\omega K}>1.5$ $\hbox{GeV}/c^2$). The hatched
histogram is from $\omega$ sidebands, and the curves in (b) show the
fit described in the text.}
\end{figure}

The Dalitz plot of events satisfying
$|m_{\pi^+\pi^-\pi^0}-m_{\omega}|<0.04$ $\hbox{GeV}/c^2$ is shown in
Fig.~\ref{mkaka}(a). There are vertical and horizontal directions
bands corresponding to $K_1(1270)^{\pm} \to \omega K^{\pm}$, and a
diagonal band corresponding to a $K^+K^-$ resonance. After requiring
$m_{\omega K}>1.5$ $\hbox{GeV}/c^2$ to remove the $K_1(1270)^{\pm}$,
the $K^+ K^-$ invariant mass distribution is shown in
Fig.~\ref{mkaka}(b). There is a cluster of events near 1.7
$\hbox{GeV}/c^2$, and there is no similar structure in $\omega$
sideband events ($0.06 < m_{\pi^+\pi^-\pi^0}-m_{\omega} < 0.1\
\hbox{GeV}/c^2\ \hbox{or} -0.1 < m_{\pi^+\pi^-\pi^0}-m_{\omega} <
-0.06 \ \hbox{GeV}/c^2) $, as shown in the hatched histogram.
Assuming the peak is $f_0(1710)$, a fit with the mass fixed at 1.714
$\hbox{GeV}/c^2$ and width fixed at 140 $\hbox{MeV}/c^2$~\cite{PDG}
gives $18.9 \pm 6.2$ events.  The statistical significance for
$\psip \to \omega f_0(1710)$ is $3.7\sigma$, and the detection
efficiency for this decay mode is $(2.62 \pm 0.06)\%$. No signal for
$e^+e^-\to \omega f_0(1710)$ is observed  at $\sqrt{s}=3.65$ GeV.

\begin{figure}[htbp]
\centerline{\psfig{file=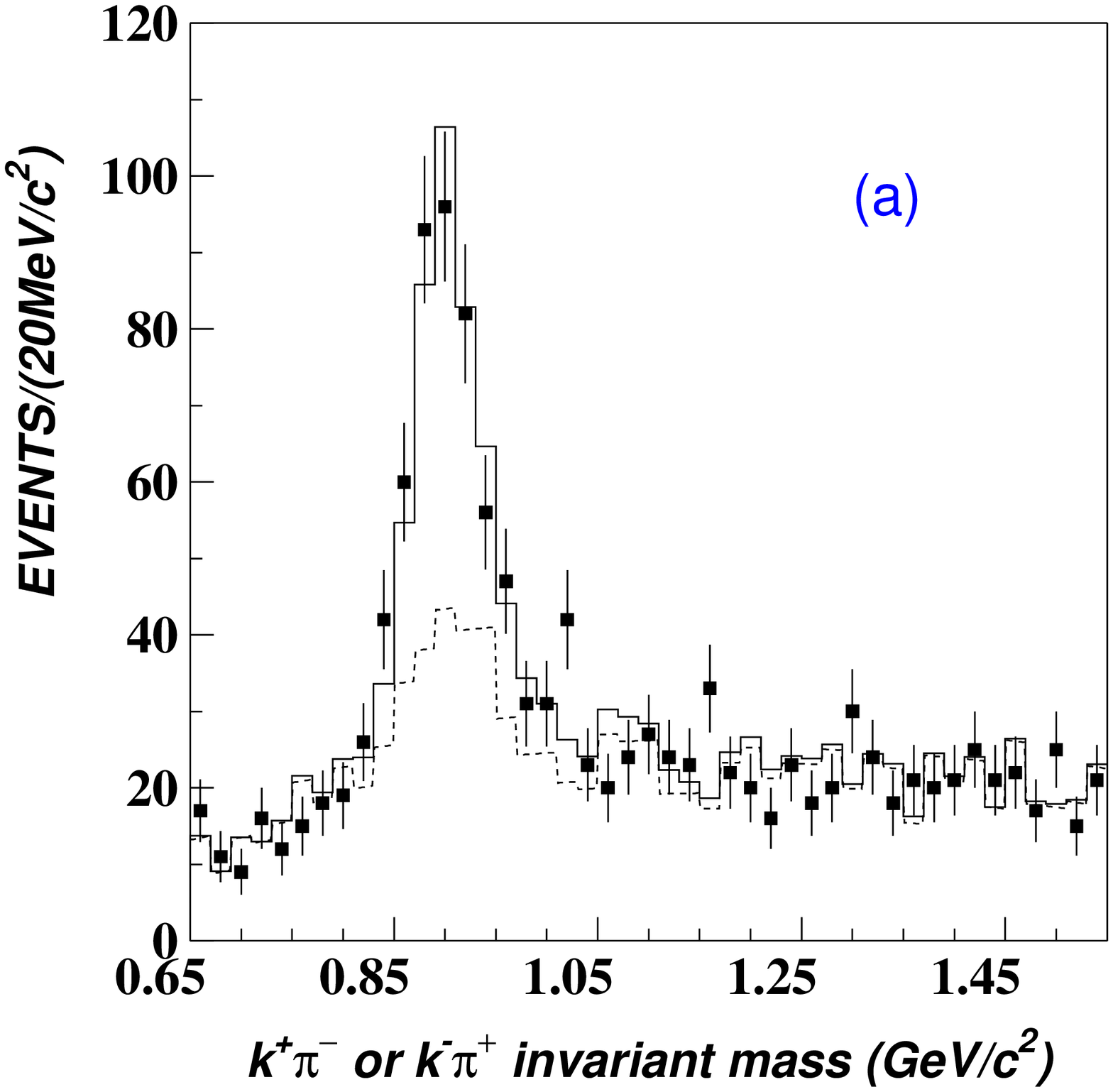,width=7.cm,height=7.cm}
            \psfig{file=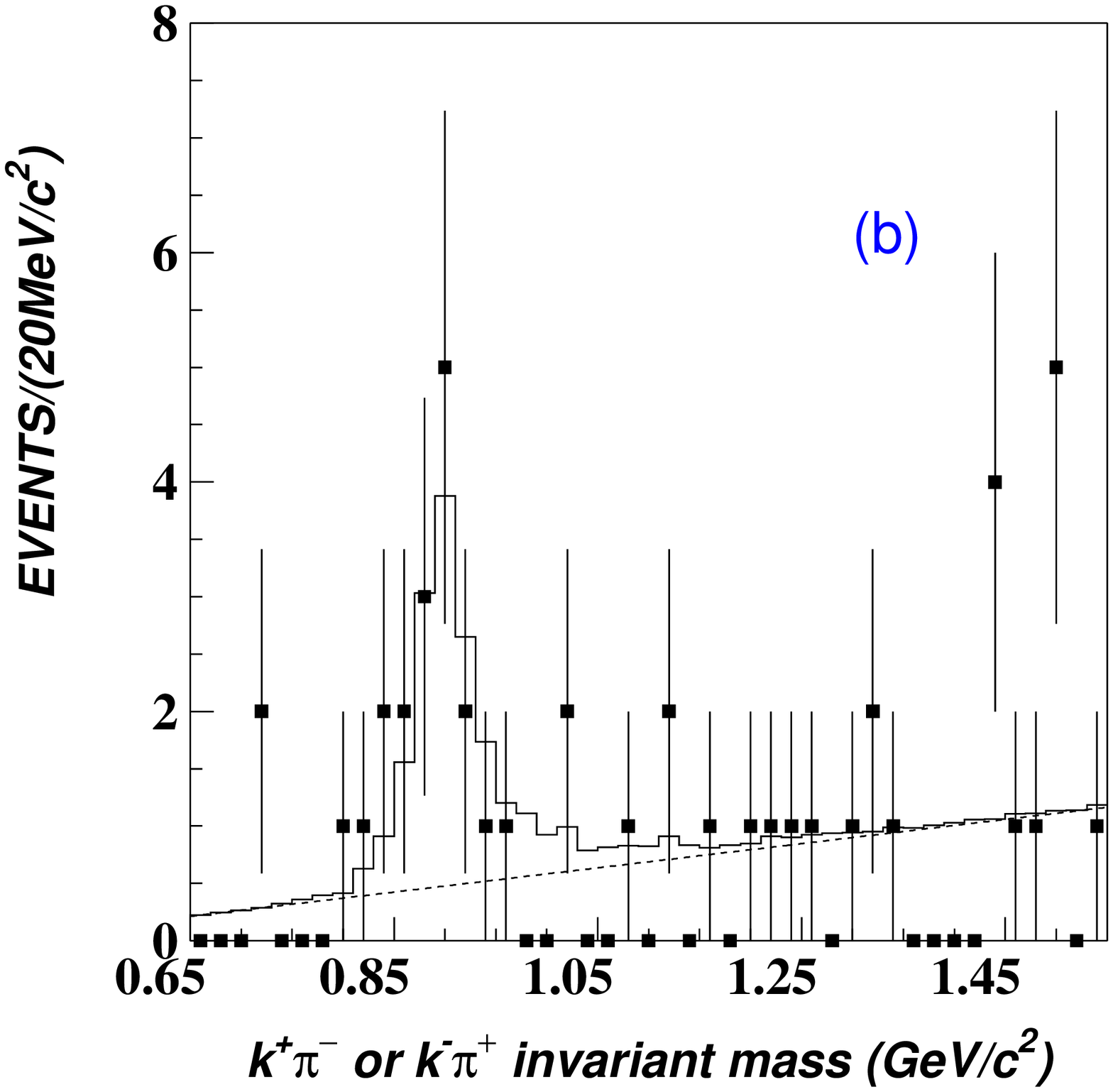,width=7.cm,height=7.cm}}
                    \caption{\label{kstar} $K^{\pm}\pi^{\mp}$
                    invariant mass distributions for $K^+ K^- \pi^+ \pi^-
\pi^0$ candidate events from (a) $\sqrt{s}=3.686$
GeV data and (b) $\sqrt{s}=3.65$ GeV data.
The squares with error bars are data,
the histograms are the fit, and the dashed lines are the sum of the
$K \pi$ mass distribution from $\pi^0$ sideband events and a Legendre polynomial.}
\end{figure}

\begin{figure}[htbp]
\centerline{\psfig{file=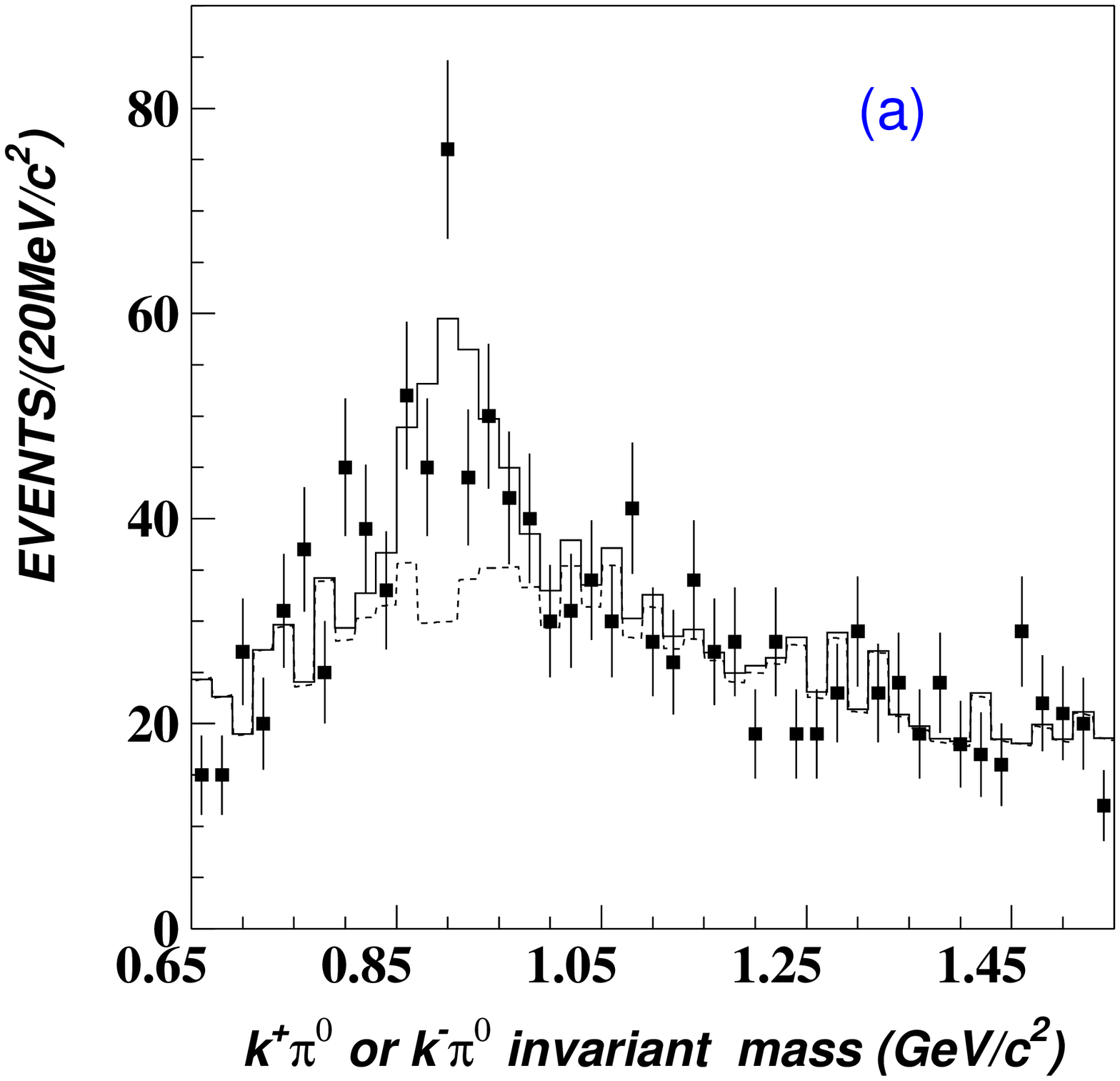,width=7.cm,height=7.cm}
            \psfig{file=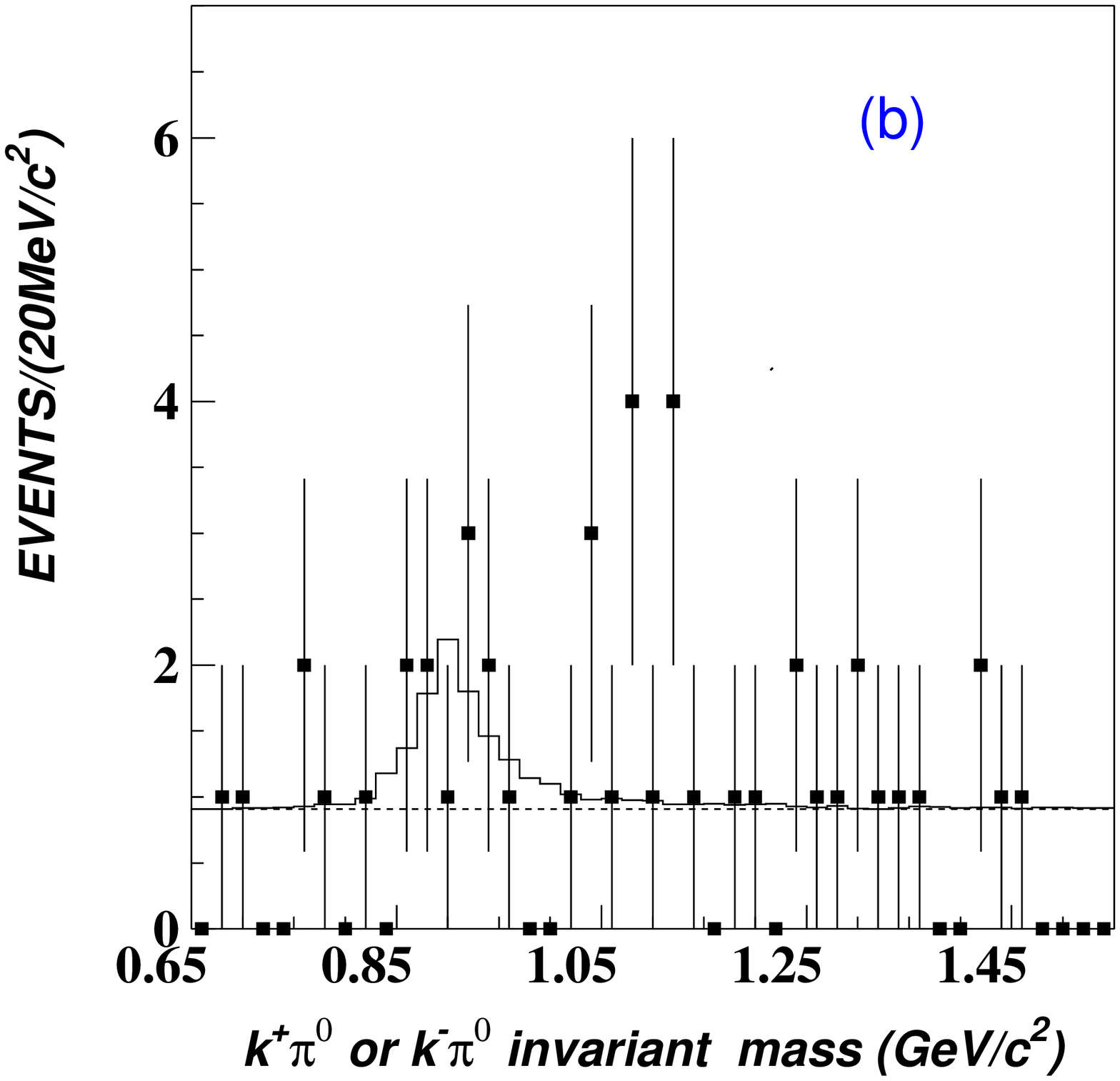,width=7.cm,height=7.cm}}
             \caption{\label{kstarpm} $K^{\pm}\pi^0$ invariant mass
          distributions  for $K^+ K^-\pi^+\pi^- \pi^0$
          candidate events from (a) $\sqrt{s}=3.686$ GeV data and (b)
          $\sqrt{s}=3.65$ GeV data. The squares with error bars are data,
          the histograms are the fit, and the dashed lines are the sum
          of the $K \pi$ mass distribution from $\pi^0$ sideband events and a
          Legendre polynomial.}
 \end{figure}

After requiring $|m_{\pi^+\pi^-\pi^0}-0.783|>0.04$ $\hbox{GeV}/c^2$
to remove $\psip \to \omega K^+ K^-$ events, the $K^{\pm} \pi^{\mp}$
mass distributions of $K^+K^-\pi^+\pi^-\pi^0$ candidates are shown
in Fig.~\ref{kstar} for (a) $\sqrt{s}=3.686$ GeV data and (b)
$\sqrt{s}=3.65$ GeV data.  By fitting the $K \pi$ invariant mass
spectrum with the Monte Carlo determined shape for signal, the $K
\pi$ mass distribution of $\pi^0$ sideband events ($0.18 < m_{\gamma
\gamma} < 0.21\ \hbox{GeV}/c^2\ \hbox{or}\ 0.06 < m_{\gamma \gamma}
< 0.09\ \hbox{GeV}/c^2 $) to describe the peaking background (mainly
$\gamma \kstaro K^- \pi^+ +c.c.)$, and a Legendre polynomial for
other backgrounds, as shown in Fig.~\ref{kstar}(a), $281\pm30$
events at $\sqrt{s}=3.686$ GeV for $\kstaro K^- \pi^+ \pi^0 +c.c.$
are obtained. For the continuum data, the peaking background is
negligible, and the fit is performed with a signal shape and a
Legendre background polynomial and yields $15 \pm 5$ events, as
shown in Fig.~\ref{kstar}(b). Similarly, the $K^{\pm} \pi^0$ mass
distributions are shown in Fig.~\ref{kstarpm}, and $150\pm26$ events
at $\sqrt{s}=3.686$ GeV and $6 \pm 5$ events at $\sqrt{s}=3.65$ GeV
are obtained for $\kstarp K^- \pi^+ \pi^- +c.c.$, as shown in
Figs.~\ref{kstarpm}(a) and (b). After subtracting the continuum
contributions, we obtain $238 \pm 34$ events for $\psip\to \kstaro
K^- \pi^+ \pi^0 +c.c.$ with a detection efficiency of $(3.00\pm
0.06)\%$ and $133\pm 30$ events for $\psip\to \kstarp K^- \pi^+
\pi^- + c.c.$ with a detection efficiency of $(3.02\pm0.06)\%$.

\begin{figure}[htbp]
\centerline{\psfig{file=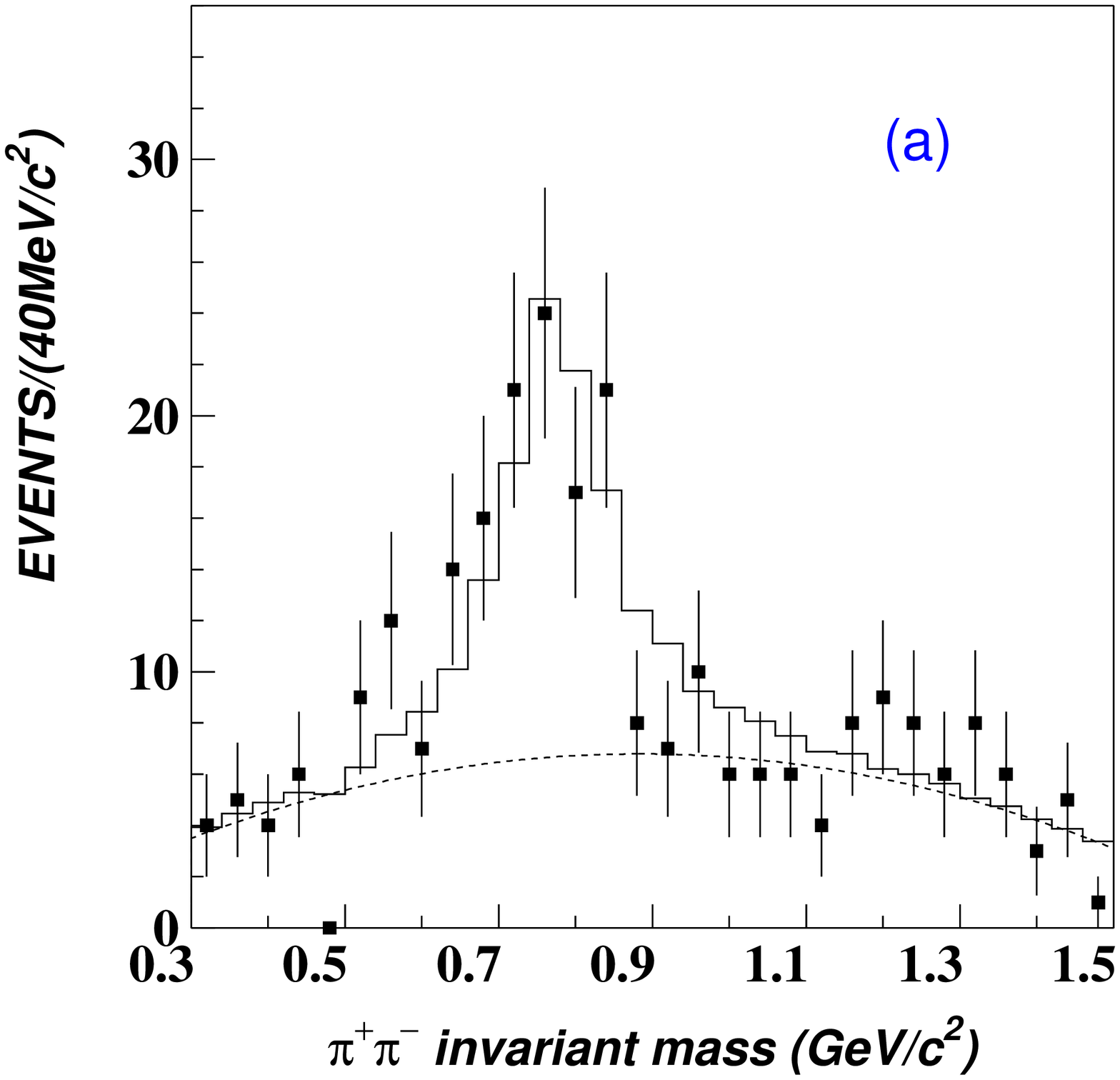,width=7.cm,height=7.cm}
            \psfig{file=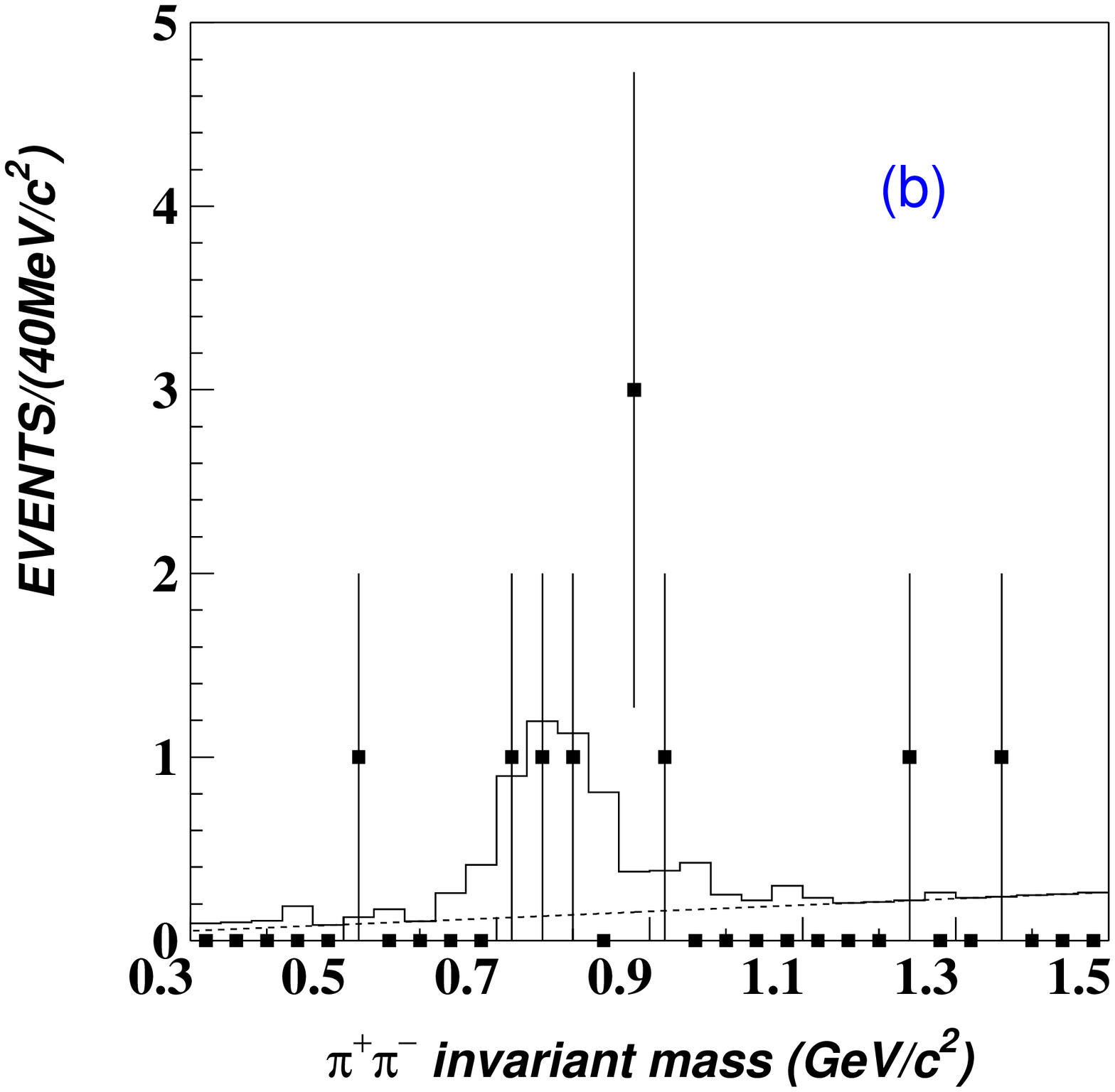,width=7.cm,height=7.cm}}
        \caption{\label{rou}
        $\pi^+\pi^-$ invariant mass
        distributions from (a) $\sqrt{s}=3.686$ GeV data
         and (b) $\sqrt{s}=3.65$ GeV data for $\kstarp K^- \pi^+ \pi^- + c.c.$
candidate events. The squares with error bars are data, the
histograms are the fit, and the dashed curves are background shapes
from the fit.}
\end{figure}

\begin{figure}[htbp]
\centerline{\psfig{file=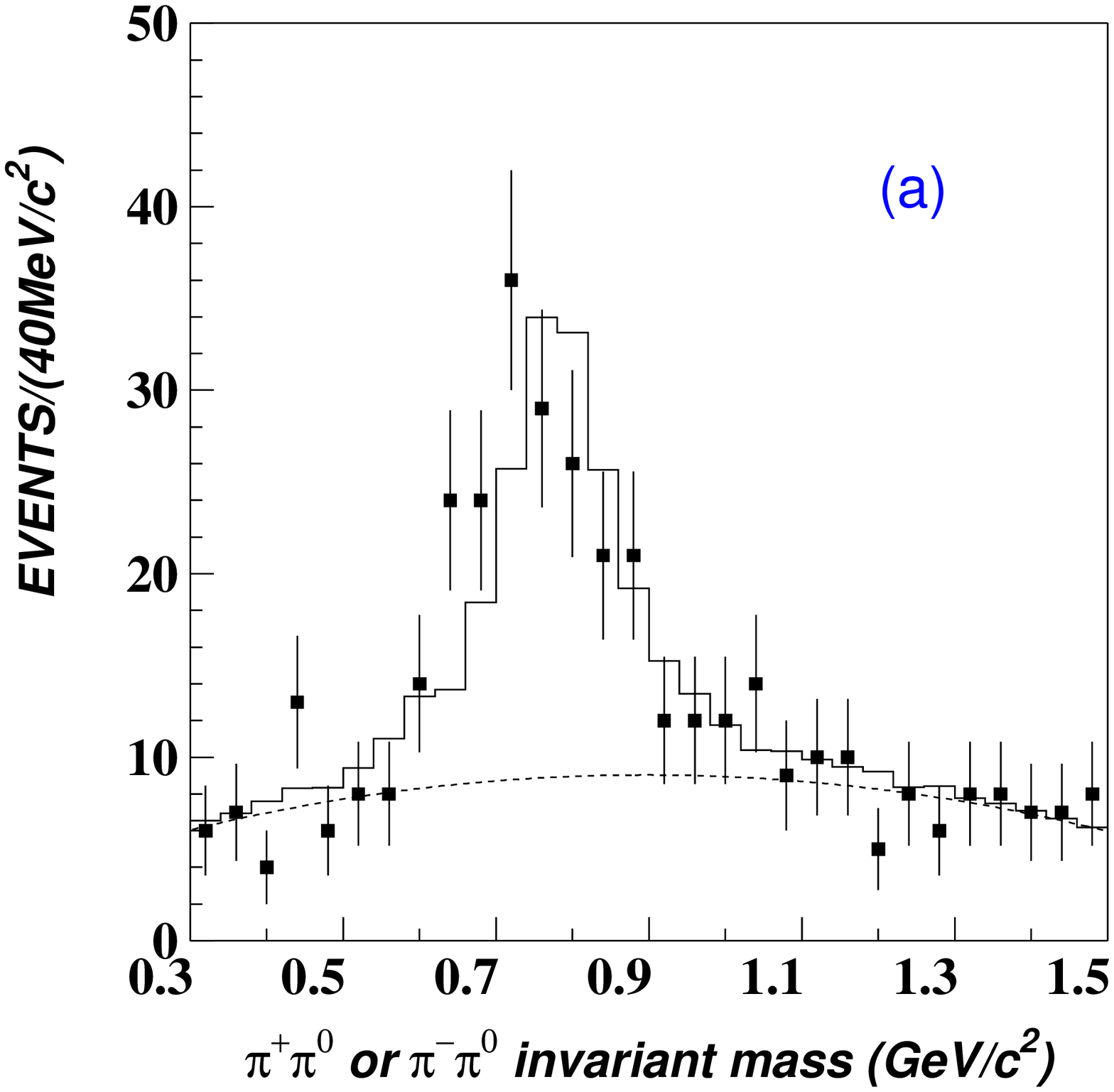,width=7.cm,height=7.cm}
            \psfig{file=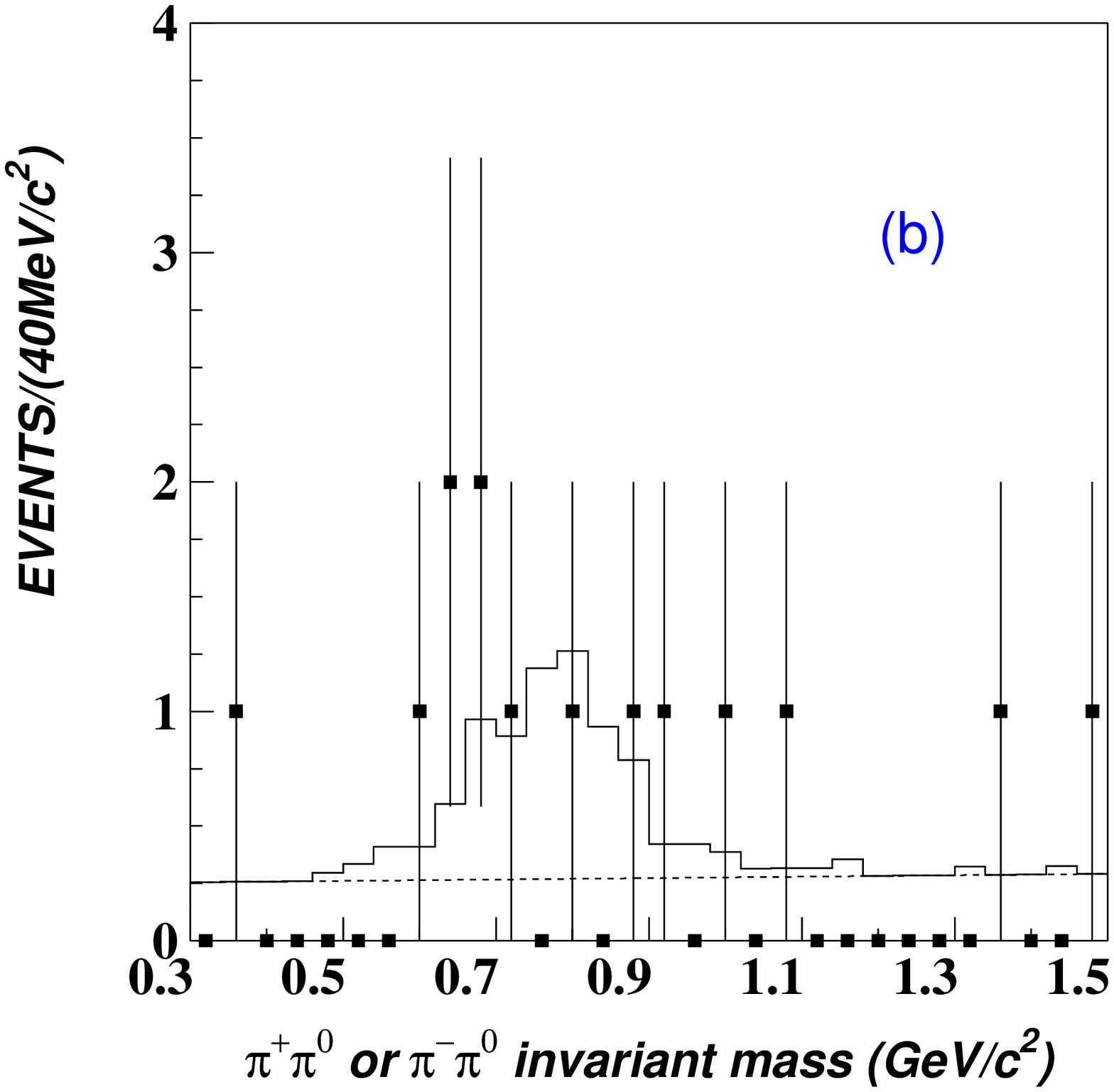,width=7.cm,height=7.cm}}
                    \caption{\label{roupm} $\pi^{\pm}\pi^0$ invariant mass
            distributions from (a) $\sqrt{s}=3.686$ GeV data
            and (b) $\sqrt{s}=3.65$ GeV data for $\kstaro K^- \pi^+\pi^0 + c.c.$
candidate events. The squares with error bars are data, the
histograms are the fit, and the dashed curves are background shapes
from the fit.}
\end{figure}

The $\pi^+\pi^-$ invariant mass distributions after requiring
$|m_{K^{\pm} \pi^0}-0.896|<0.06$ $\hbox{GeV}/c^2$ are shown in
Fig.~\ref{rou}.  The spectra are fitted with Monte Carlo determined
$\rho$ shapes and Legendre background polynomials, as shown in
Figs.~\ref{rou}(a) and (b), and the fit yields
$92\pm19$ events at $\sqrt{s}=3.686$ GeV and $5 \pm 4$ events at
$\sqrt{s}=3.65$ GeV for $\kstarp K^- \rho^0 + c.c.$.
Similarly the $\pi^{\pm}\pi^0$ invariant
mass distributions are shown in Fig.~\ref{roupm} for the $\sqrt{s}=3.686$
GeV and $\sqrt{s}=3.65$ GeV data samples. The fits, shown in Figs.~\ref{roupm}(a)
and (b), yield $142\pm23$
events at $\sqrt{s}=3.686$ GeV and $6 \pm 4$ events at $\sqrt{s}=3.65$
GeV for $\kstaro K^- \rho^+ +c.c.$.
After subtracting the continuum contributions, we obtain $78
\pm 23$ $\psip\to \kstarp K^- \rho^0 +c.c.$ events with a detection
efficiency of $(2.32\pm 0.05)\%$ and $125\pm 26$ $\psip\to \kstaro K^-
\rho^+ + c.c$ events with a detection efficiency of $(2.24\pm0.05)\%$.

\begin{figure}[htbp]
\centerline{\psfig{file=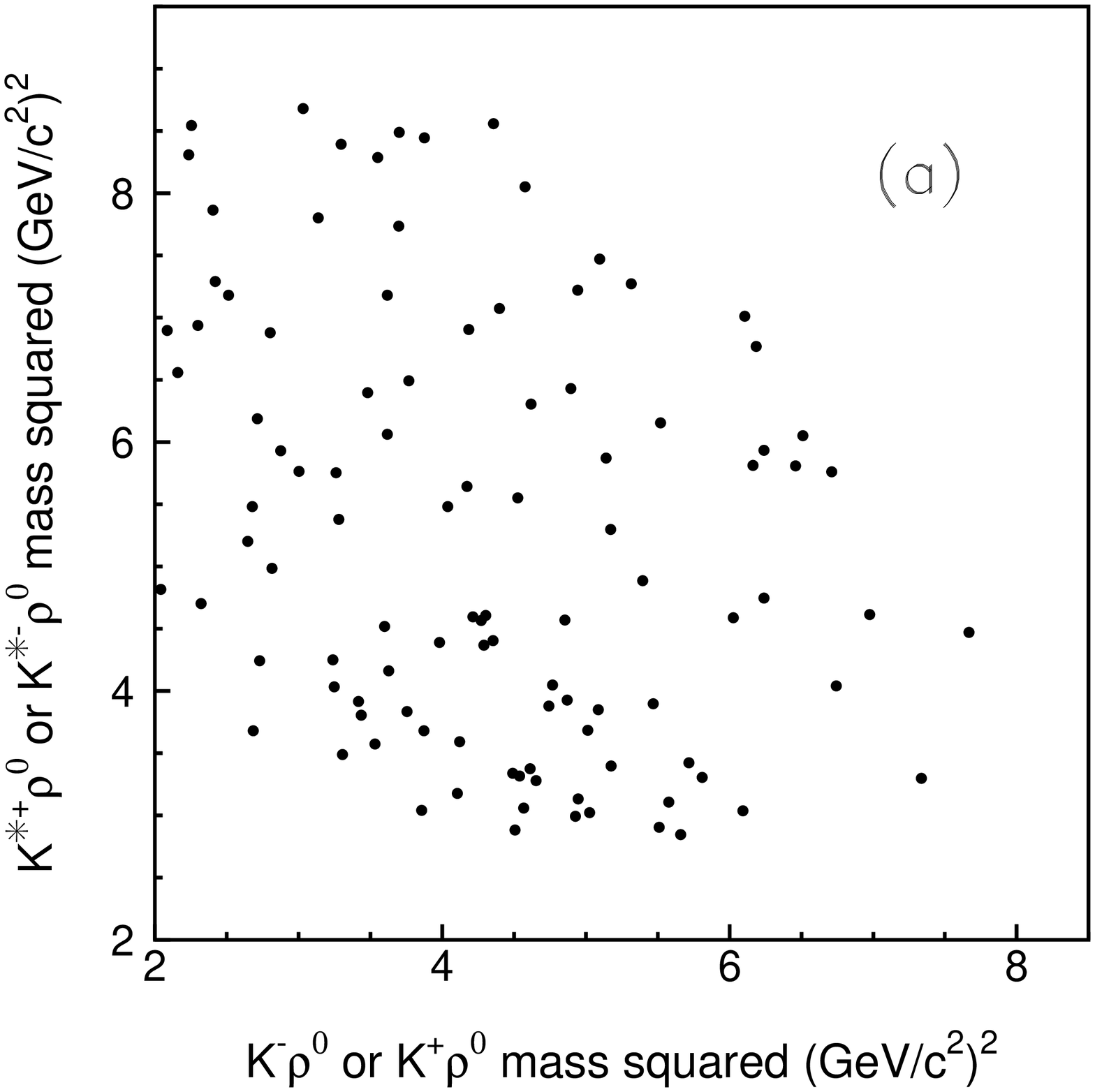,width=7.cm,height=7.cm}
            \psfig{file=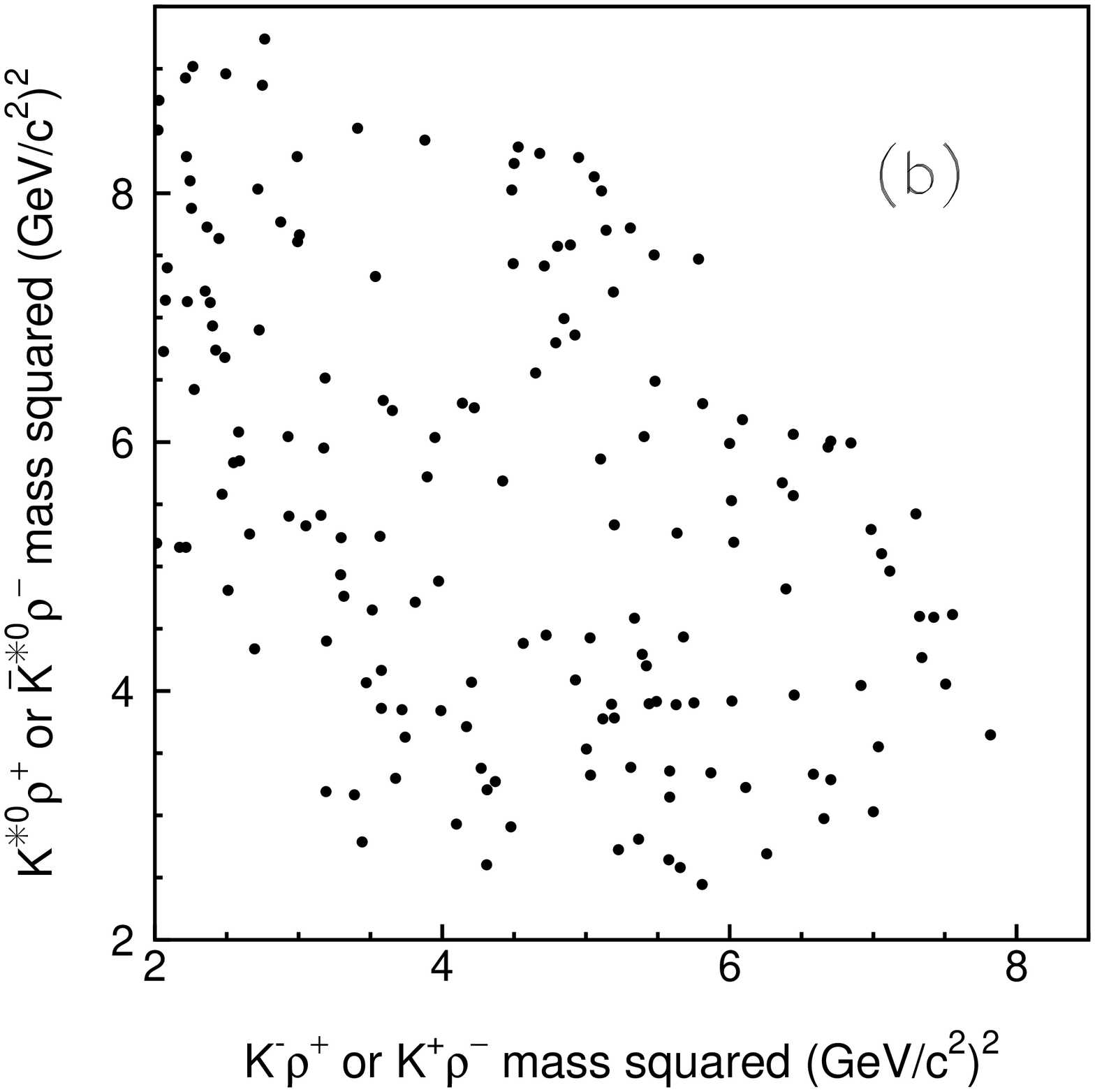,width=7.cm,height=7.cm}}
 \caption{\label{kstar-rou}  Dalitz plots of (a) $\psip \to \kstarp K^- \rho^0 +c.c.$ and (b) $\psip \to
 \kstaro K^- \rho^+ +c.c.$.
  }
\end{figure}

After requiring $|m_{K\pi}-0.896|<0.06$ $\hbox{GeV}/c^2$ and
$|m_{\pi\pi}-m_{\rho}|<0.15$ $\hbox{GeV}/c^2$, the Dalitz plots of
$\psip \to \kstarp K^- \rho^0 +c.c.$ and $\psip \to \kstaro K^-
\rho^+ +c.c.$ candidates are shown in Figs.~\ref{kstar-rou}(a) and (b),
where no further clear structures are observed.

\section{Systematic errors}
Systematic errors on the branching fractions mainly originate from
the MC statistics, the track error matrix, the kinematic fit,
particle identification, the photon efficiency, the uncertainty of
the branching fractions of the intermediate states (from
PDG)~\cite{PDG}, the $K^{\ast}(892)$ simulation, the fitting, and
the total number of $\psip$ events.

\begin{enumerate}
\item The MDC tracking efficiency was measured using clean channels
like $J/\psi \to \Lambda \bar{\Lambda}$ and $\psip \to \pi^+\pi^-
J/\psi, J/\psi \to \mu^+ \mu^-$. It is found that the MC
simulation agrees with data within $(1-2)\%$ for each charged track.
Therefore, $8\%$ is taken as the systematic error for events with four
charged tracks.

\item The photon detection efficiency was studied using different
methods with $J/\psi \to \pi^+\pi^-\pi^0$ events~\cite{simbes}, and
the difference between data and MC simulation is about $2\%$ for
each photon. The systematic error due to the differences between
data and Monte Carlo simulation of fake photons and the
reconstruction of the $\pi^0$ is less than 2.5\%~\cite{xinbo}. We
take $5\%$ as the systematic error for channels with two photons.

\item The systematic error associated with the kinematic fit is caused
  by differences between the momenta and error matrices from track
  fitting of the charged tracks and the energies and the directions of
  the neutral tracks for data and Monte Carlo data.  For the channels
  analyzed, $Prob(\chi^2,4)>0.01$ is required, and we take $4\%$ as
  the systematic error from the kinematic fit~\cite{VT}.
  Fig.~\ref{chisq} shows the comparison of the $\chi^2$ distributions
  for data and MC sample, where we require $|m_{\gamma
    \gamma}-0.135|<0.03$ $\hbox{GeV}/c^2$ and subtract $\pi^0$
  sideband background. MC simulation agrees with data within
  large statistical uncertainty of the data sample.

\begin{figure}[htbp]
\centerline{
\includegraphics [height=7cm,width=7cm]{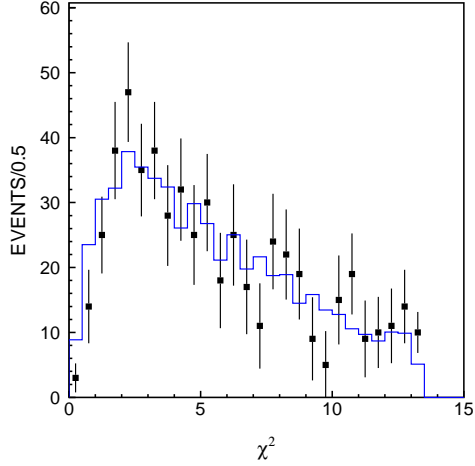}}
\caption{The $\chi^2$ distribution from the kinematic fit. The
squares with error bars are data, and the histogram is MC simulated
$\psip\to \kkppp$.} \label{chisq}
\end{figure}

\item The background uncertainties in the $K^+ K^- \pi^+ \pi^- \pi^0$,
 $\omega K^+ K^-$, $\omega f_0(1710)$, $\kstaro K^- \pi^+ \pi^0 +
 c.c.$, $\kstarp K^- \pi^+ \pi^- + c.c.$, $\kstarp K^- \rho^0 + c.c.$,
 and $\kstaro K^- \rho^+ +c.c.$ channels are estimated to be about
 $1.5\%$, $1.6\%$, $1.6\%$, $7.1\%$, $9.5\%$, $11.4\%$, and $11.3\%$,
 respectively, by changing the order of the polynomial and the fitting
 range used.  Varying the mass and width of the $f_0(1710)$~\cite{dongly}
 in the fit yields a change in the number of $\omega f_0(1710)$ events
 by $8.6\%$. Together with the background uncertainty, we take
 $8.8\%$ as the systematic error for $\psip \to \omega f_0(1710)$ due
 to fitting.

\item Pure $\pi$ and $K$ samples were selected, and the particle identification
efficiency was measured as a function of the track momentum. On the average,
a $1.3\%$ efficiency difference per $\pi$ track and a $1.0\%$ difference
per $K$ track are observed between data and MC simulation. The total
systematic error for $K^+K^-\pi^+\pi^-\pi^0$ is taken as $4.6\%$.

\item The $K^{\ast}(892)$ is simulated with a P-wave
 relativistic Breit-Wigner function, with the width
  $\Gamma=\Gamma_0 \frac{m_0}{m} \frac{1+r^{2}p_0^2}{1+r^{2}
   p^2}\Big[\frac{p}{p_0}\Big]^3$, where $r$ is the interaction radius
and the value $(3.4\pm0.6\pm0.3)$ $\hbox{(GeV/c)}^{-1}$ measured by
a $K^-\pi^+$ scattering experiment \cite{r} is used as an estimation of
the interaction radius.
Varying the value of $r$ by $1\sigma$, the detection efficiencies
for $\psip$ decaying to $\kstaro K^- \pi^+ \pi^0 +c.c.$, $\kstarp
K^- \pi^+ \pi^-+c.c.$, $\kstarp K^- \rho^0 + c.c.$, and $\kstaro K^-
\rho^+ +c.c.$ changed by $3.0\%$ for the first
two modes and $7.0\%$ for the last two modes, which are taken
 as systematic errors for the uncertainty of the $r$ value.

\item For the $\kstaro K^- \pi^+ \pi^0 +c.c.$ decay mode, there are
  backgrounds that also peak in the signal region of the $K \pi$ mass plot.
  The largest is from $\gamma \kstaro K^- \pi^+ +c.c.$ events combined
  with fake photons.  We use $\pi^0$ sideband events to estimate these
  background contributions. In the analysis, the $\pi^0$ sideband
  range is from 3 to 5 $\sigma$, where $\sigma$ is 15 MeV/c$^2$.
  If the $\pi^0$ sideband range is chosen between 5 and 7 $\sigma$,
  the difference with the standard sideband is $8.0\%$. If the $\pi^0$
  sideband range is from 2 to 3 $\sigma$, the difference is
  $13.7\%$. So we take $13.7\%$ as the systematic error due to
  the $\pi^0$ sideband definition in the fitting. Using a similar
  procedure for $\kstarp K^- \pi^+ \pi^- +c.c.$ decay, $8.1\%$ is
  determined as the systematic error.

\item In calculating the Born order cross section for $e^+ e^- \to K^+ K^-
\pi^+ \pi^- \pi^0$, the error for the integrated luminosity is
$4\%$. The other systematic errors are similar to those for $\psip
\to K^+ K^- \pi^+ \pi^- \pi^0$. The total systematic error for
$\sigma(e^+ e^- \to K^+ K^- \pi^+ \pi^- \pi^0)$ is $12.1\%$.
\end{enumerate}

Table~\ref{sumerror} lists all the systematic errors from different
sources, the total systematic errors for $\psi(2S) \to K^+ K^- \pi^+
\pi^- \pi^0$, $\psi(2S) \to \omega K^+ K^- $, $\psi(2S) \to \omega
f_0(1710)$,  $\psi(2S) \to K^{\ast0} K^- \pi^+ \pi^0+c.c.$,
$\psi(2S) \to K^{\ast+} K^- \pi^+ \pi^- +c.c.$, $\psi(2S) \to
K^{\ast+} K^- \rho^0 +c.c.$ and $\psi(2S) \to K^{\ast0} K^- \rho^+
+c.c.$ are $12.1\%$, $12.2\%$, $15.1\%$, $19.9\%$, $17.7\%$,
$18.1\%$ and $18.1\%$ respectively.

\begin{table}[hbtp]
\begin{center}
\caption {Summary of systematic errors ($\%$). Errors common to
all modes are only listed once.} \vspace{3.0mm} {\scriptsize
\begin{tabular}{l c c c c c c c}
  \hline\hline
  Source &$K^+ K^- \pi^+ \pi^- \pi^0$ & $\omega K^+ K^-$ &$\omega f_0(1710)$
               &$K^{\ast0} K^- \pi^+ \pi^0$ & $K^{\ast+} K^- \pi^+ \pi^-$
           &$K^{\ast+} K^- \rho^0$      & $K^{\ast0} K^- \rho^+ $
           \\
 &  &  &  & $+c.c.$ & $+c.c.$ &$+c.c.$ &$+c.c.$ \\\hline
MC statistics & 1.4 & 1.5 & 2.3 &2.0 &1.8 &2.2 &2.3 \\
Tracking efficiency    &     &     &     &8.0 &    &    & \\
Kinematic fit        &     &     &     &4.0 &    &    & \\
PID efficiency        &     &     &     &4.6 &    &    & \\
Photon ID and &     &     &     &  &    &    & \\
$\pi^0$ reconstruction &     &     &   &5.0 &    &    & \\
Fitting             & 1.5 & 1.6 & 8.8 &7.1  &9.5 & 11.4 &11.3\\
Branching fractions & $\cdots$ & 0.8 & 0.8 &$\cdots$  & $\cdots$&$\cdots$   &$\cdots$ \\
$r$ uncertainty       & $\cdots$ & $\cdots$ & $\cdots$ &3.0  &3.0 & 7.0 &7.0 \\
$\pi^0$ sideband   & $\cdots$ &$\cdots$ & $\cdots$ &13.7 &8.1 & $\cdots$ &$\cdots$ \\
$N_{\psip}$         &     &     &     &4.0 &    &    & \\\hline
Sum                 &12.1&12.2&15.1&19.9&17.7&18.1 &18.1     \\
\hline\hline
\end{tabular}
} \label{sumerror}
 \end{center}
\end{table}

\section{Results and discussion}

To obtain the branching fraction of $\psip \to X$, we must subtract
the contribution from the continuum process. This is estimated using
continuum data at $\sqrt{s}=3.65$ GeV, normalized by $f$:
\begin{eqnarray*}
  f=\frac{\displaystyle\mathcal{L}_{3.686}\times\sigma^{cont}_{3.686}}
    {\displaystyle\mathcal{L}_{3.650}\times\sigma^{cont}_{3.650}},
    \end{eqnarray*}
where $\displaystyle\mathcal{L}_{\sqrt{s}}$ is the integrated
luminosity at $\sqrt{s}$, and $\sigma^{cont}_{\sqrt{s}}$ is the Born
order cross section of the continuum
process at $\sqrt{s}$, which is $s$ dependent and can be expressed in terms of
a form factor $\mathcal{F}(s)$:
\begin{eqnarray}
\sigma^{cont}_{\sqrt
{s}}=\frac{4\pi\alpha^{2}}{3s}\times|\mathcal{F}(s)|^{2},
  \label{born1}
  \end{eqnarray}
where $\alpha$ is the QED fine structure constant. Assuming
$|\mathcal{F}(s)| \propto 1/s$, we get $f=2.89$. The branching
fraction of $\psip\to X$ can be calculated from
$$  \mathcal{B}[\psip \to  X] =\frac{\displaystyle
        N^{obs}_{3.686}-N^{obs}_{3.650}\times f}
            {\displaystyle \varepsilon_{\psip}\times
N^{tot}_{\psip} \times B(X \to Y)},$$
where $X$ is the intermediate state and $Y$ is the final state.

Using the numbers obtained above and listed in Table~\ref{datasum}, we get
$$B(\psip\to K^+K^-\pi^+\pi^-\pi^0)=(1.17\pm0.10\pm 0.15)\times
10^{-3},$$
  $$B(\psip\to \omega K^+ K^-)=(2.38 \pm 0.37 \pm 0.29)\times 10^{-4},$$
  $$B(\psip\to \omega f_0(1710), f_0(1710)\to K^+K^-)=(5.9 \pm 2.0 \pm
0.9)\times 10^{-5},$$
  $$B(\psip\to \kstaro K^- \pi^+\pi^0+c.c.)=(8.6 \pm 1.3 \pm
1.8)\times 10^{-4},$$
  $$B(\psip\to \kstarp K^- \pi^+\pi^-+c.c.)=(9.6 \pm 2.2 \pm
1.7)\times 10^{-4},$$
  $$B(\psip\to \kstarp K^- \rho^0+c.c.)=(7.3 \pm 2.2 \pm 1.4)\times
10^{-4},$$
  $$B(\psip\to\kstaro K^-\rho^+ + c.c.)=(6.1 \pm 1.3 \pm
1.2)\times 10^{-4},$$
 where the first errors are statistical and the second are systematic.
The measured $\psip \to K^+K^-\pi^+\pi^-\pi^0$ branching fraction
agrees well with the value of  $(1.27\pm0.05\pm0.10)\times10^{-3}$
obtained by CLEO~\cite{CLEO}, and the $\psip \to \omega K^+ K^-$
branching fraction agrees with BESI~\cite{omegakk} and CLEO results
within 1.5$\sigma$. The other five modes are first observations.

The Born order cross section for $e^+ e^- \to K^+ K^-\pi^+\pi^-\pi^0$
at $\sqrt{s}=3.65$ GeV is
    $$\sigma^{B}_{e^+ e^- \to K^+ K^-\pi^+\pi^-\pi^0}=\frac{N^{obs}}{
    \mathcal{L} \epsilon (1+\delta)}=(171 \pm 35 \pm 21)~\mbox{pb},$$
 where $\epsilon$ is the detection efficiency
 obtained from the MC simulation, $\mathcal{L}$ is the integrated luminosity and
 $1+\delta$ is the radiative correction factor which is
 1.23~\cite{G}.

\begin{table}[htbp]
\begin{center}
\caption{\label{datasum} Numbers used in the branching fraction
calculations. The number of events due to $\psip$ decay,
$N_{\psip}^{obs}$, is computed according to $N_{3.686}^{obs}-f\times
N_{3.65}^{obs}$. } \vspace{3.0mm}
 {\scriptsize \begin{tabular}{c c c c c c c c }
   \hline\hline
 Quantity & $K^+ K^- \pi^+ \pi^-\pi^0$ &$ \omega K^+ K^- $& $\omega f_0(1710)$
          & $K^{\ast 0} K^- \pi^+\pi^0$& $K^{\ast +} K^- \pi^+\pi^- $
          &$K^{\ast  +} K^- \rho^0$  & $K^{\ast 0} K^- \rho^+ $ \\
          &    &   &  &$+c.c.$&$+c.c.$&$+c.c.$&$+c.c.$   \\\hline
  $ N_{3.65}^{obs}$    & $35\pm7$   & $0^{+1.3}_{-0}$  & -- &$15\pm5$
&$6\pm5$&$5\pm4$&$6\pm4$   \\
$N_{3.686}^{obs}$ & $698\pm41$ & $78\pm11$ & $18.9\pm6.2$
&$281\pm30$ &$150\pm26$&$92\pm19$ &$142\pm23$ \\
  $N_{\psip}^{obs}$    & $597\pm46$ & $78^{+12}_{-11}$ &
$18.9\pm6.2$ &$238\pm34$ &$133\pm30$&$78\pm23$ &$125\pm26$ \\
  $\epsilon (\%)$  & $3.68\pm0.05$ &$2.66\pm 0.04$
&$2.62\pm0.06$&$3.00\pm0.06$&$3.02\pm0.06$ &$2.32\pm0.05$&$2.24\pm0.05$  \\
  $N_{\psip}(10^6)$ & &  & & $14(1\pm4\%)$ &  & &  \\
\hline\hline
 \end{tabular}
    }
\end{center}
\end{table}

  To test the $12\%$ rule, we also list in Table~\ref{12}
  the ratio $Q_h$ of the $\psip$ and $J/\psi$ branching fractions for
  the three channels $\psip \to K^+K^-\pi^+\pi^-\pi^0$, $\psi(2S)\to
  \omega K^+ K^-$, and $\psi(2S)\to \omega f_0(1710), f_0(1710)\to
  K^+K^-$.
  $B(J/\psi \to  K^+ K^-\pi^+\pi^-\pi^0)$ is taken from the
PDG~\cite{PDG}, while the values on $B(J/\psi \to \omega K^+ K^-)$
and $B(J/\psi \to\omega f_0(1710))$ come from Ref.~\cite{dongly}, a
recent measurement using a partial
wave analysis and the  BESII $J/\psi$ sample.
All three modes obey the $12\%$
rule within $1\sigma$. The four modes with $K^{\ast}$ we observed in
$\psip$ decays are not measured in $J/\psi$ decays, so $Q_h$ values
can not be computed.

\begin{table}[htbp]
  \caption{\label{12} Branching fractions for $\psip$ and $J/\psi$
           hadronic decays and $Q_h$ values. The errors are the
           quadratic sum of the statistical and systematic errors.}
  \vspace{3mm}
  {\small
  \begin{tabular}{c c c c}
  \hline\hline
  Channel  &$B_{\psip\to h}(10^{-4})$ &$B_{J/\psi \to h}(10^{-4})$&$Q_h(\%)$\\\hline
$K^+K^-\pi^+\pi^-\pi^0$&$11.7\pm1.8$ &$120\pm28$~\cite{PDG}&$9.8\pm2.8$\\
$\omega K^+K^-$ & $2.38\pm0.47$&$16.8\pm2.1$~\cite{dongly}&$14.2\pm3.4$  \\
$\omega f_0(1710)\to \omega K^+ K^-$&
$0.59\pm0.22$&$6.6\pm1.3$~\cite{dongly}&$8.9\pm3.8$  \\\hline\hline
\end{tabular}
}
\end{table}

\section{Acknowledgment}

The BES collaboration thanks the staff of BEPC for their hard
efforts. This work is supported in part by the National Natural
Science Foundation of China under contracts Nos. 10491300, 10225524,
10225525, 10425523, the Chinese Academy of Sciences under contract
No. KJ 95T-03, the 100 Talents Program of CAS under Contract Nos.
U-11, U-24, U-25, and the Knowledge Innovation Project of CAS under
Contract Nos. U-602, U-34 (IHEP), the National Natural Science
Foundation of China under Contract No. 10225522 (Tsinghua
University), and the Department of Energy under Contract
No.DE-FG02-04ER41291 (U Hawaii).


\begin{thebibliography}{99}
\bibitem{dongly}BES Collab., J.~Z.~Bai {\itshape et
  al}., Phys. Lett. B {\bf603}, 138 (2004). The branching fraction of
$B(J/\psi \to \omega K^+K^-)$ is calculated using the numbers in
this paper, the total error for $\psip \to \omega f_0(1710)$ is
$19.7\%$ and the $f_0(1710)$ intensity is $(38\pm6)\%$ of the data,
where the error is almost entirely systematic. So we estimated the
systematic error for $J/\psi \to \omega K^+ K^-$ is $11.8\%$.
\end{thebibliography}
\end{document}